# Domain wall resistance in CoFeB-based heterostructures with interface Dzyaloshinskii-Moriya interaction

Yuto Ishikuro[1], Masashi Kawaguchi[1], Yong-Chang Lau[1,2], Yoshinobu Nakatani[3] and Masamitsu Hayashi[1,2]*

[1]*Department of Physics, The University of Tokyo, Bunkyo, Tokyo 113-0033, Japan*

[2]*National Institute for Materials Science, Tsukuba 305-0047, Japan*

[3]*University of Electro-Communications, Chofu, Tokyo 182-8585, Japan*

We have studied the domain wall resistance in W/Ta/CoFeB/MgO heterostructures. The Ta layer thickness is varied to control the type of domain walls via changes in the interfacial Dzyaloshinskii Moriya interaction. We find a nearly constant domain wall resistance against the Ta layer thickness. Adding contributions from the anisotropic magnetoresistance, spin Hall magnetoresistance and anomalous Hall effect describe well the domain wall resistance of the thick Ta layer films. However, a discrepancy remains for the thin Ta layer films wherein chiral Néel-like domain walls are found. These results show the difficulty of studying the domain wall type from resistance measurements.

*Email: hayashi@phys.s.u-tokyo.ac.jp

The Dzyaloshinskii−Moriya interaction (DMI)[1,2] at the interface between a heavy metal (HM) and a ferromagnetic metal (FM) layers[3] enables stabilization of homochiral Néel domain walls (DWs)[4]. Néel DWs can be driven by current with high velocity via the spin Hall effect of the HM layer[5-8]. Significant effort has been made to develop means to determine the character of the DWs, whether they form a Néel or Bloch walls, and the strength of the DMI[9-14]. It has been demonstrated previously that the anisotropic magnetoresistance (AMR) of the FM layer can be used to identify the structure and type of DWs[15,16]. Direct determination of the DW type using resistance measurements allows simple evaluation of the DMI, including its spatial distribution. Local changes in the DMI at the HM/FM interface has been reported to be non-negligible[17,18].

Here we study DW resistance in Sub/W/Ta/CoFeB/MgO heterostructures. A thin Ta layer is inserted between the W and CoFeB layers to vary the strength of the DMI. We compare the DW resistance with calculations that includes contributions from the AMR, the spin Hall magnetoresistance (SMR)[19-22] and the anomalous Hall effect (AHE). We find that such estimation agrees well with the DW resistance for the thick Ta layer films; however, a discrepancy remains for the thinner Ta layer films. Other contribution to the DW resistance, including the recently discovered chiral DW resistance[23], may account for the discrepancy. Despite the large SMR, these results show the difficulty of evaluating the type of DW using simple resistance measurements.

Films are grown on Si substrates using magnetron sputtering. Figure 1(a) shows schematic illustration of the film stack used here, i.e. Sub./3 W/$d$ Ta/1 CoFeB/2 MgO/1 Ta (units in nanometer). The thickness of the Ta insertion layer $d$ is varied to control the size of the DMI. All films are annealed at 300 degree C for 1 hour. Optical lithography and Ar ion milling are used to form wires and Hall bars. The width and length of the wires are typically ~5 μm and ~30 μm, respectively. The width ($w$) and the distance ($L$) between the longitudinal voltage probes of the Hall bars are ~10 μm and ~25 μm, respectively. Figure 1(b) shows a Kerr microscopy images of typical wire and Hall bar with the definition of the coordinate axis. Voltage pulses are applied to the wire to nucleate DWs. The current associated with the pulse flows along the $x$ direction. External magnetic fields are applied along the $x$, $y$, and $z$ directions, which we refer to as $H_x$, $H_y$, and $H_z$, respectively.

Figures 1(c) and 1(d) show the Ta layer thickness dependence of the saturation magnetization per unit volume $M_s$ and the magnetic anisotropy energy density $K_{\text{eff}}$, respectively. $M_s$ and $K_{\text{eff}}$ are evaluated with vibrating sample magnetometry (VSM) using unpatterned films. Both $M_s$ and $K_{\text{eff}}$ decrease as $d$ is increased. We infer that intermixing of Ta and CoFeB at the interface causes such variation[24].

The transport properties of the films are summarized in Figs. 1(e) and 1(f). Figure 1(e) shows the Ta layer thickness dependence of the anomalous Hall angle $\varphi_{\text{AHE}}^{\text{eff}}$ measured using the Hall bars. The anomalous Hall resistance $\Delta R_{xy}$, obtained from the difference of the Hall resistance $R_{xy}$ when the magnetization of the CoFeB layer points along $+z$ and $-z$, is divided by the longitudinal resistance $R_{xx}$ and multiplied by a geometrical factor ($L/w$) to estimate $\varphi_{\text{AHE}}^{\text{eff}}$, i.e. $\tan(\varphi_{\text{AHE}}^{\text{eff}}) = \frac{1}{2}\frac{\Delta R_{xy}}{R_{xx}}\frac{L}{w}$. $\varphi_{\text{AHE}}^{\text{eff}}$ decreases with increasing $d$ due to current shunting into the Ta layer. The Ta layer thickness dependence of the AMR and the SMR are shown in Fig. 1(f). The longitudinal resistance of the Hall bars measured against magnetic field directed along the $x$, $y$, and $z$ axes are defined as $R_{XX}^X$, $R_{XX}^Y$, and $R_{XX}^Z$, respectively. Setting $\Delta R_{XX}^X = R_{XX}^X - R_{XX}^Z$ and $\Delta R_{XX}^Y = R_{XX}^Y - R_{XX}^Z$, the magnetoresistance ratios $r_{\text{AMR}} = \Delta R_{XX}^X / R_{XX}^Z$ due to AMR and $r_{\text{SMR}} = \Delta R_{XX}^Y / R_{XX}^Z$ caused by SMR are plotted as a function of $d$ in Fig. 1(f). The AMR of the heterostructures is negligible compared to the SMR[22]. SMR decreases with increasing Ta layer thickness mostly due to current shunting effect (into Ta).

The DMI is measured using the in-plane field dependence of the current induced anomalous Hall loop shift[14] and the DW velocity[6,7,25]. For the former, $R_{xy}$ vs. $H_z$ is measured under application of a DC bias current $I_{DC}$. Figure 2(a) shows representative $R_{xy}$-$H_z$ loops for a device with $d$~0.1nm ($H_x$= -1100 Oe and $I_{DC}=\pm$6mA). When a positive (negative) current is applied, the center of the hysteresis shifts to positive (negative) $H_z$. The shift of the loop center is defined as $H_{eff}^z$. The inset to Fig. 2(b) shows $H_{eff}^z$ as a function of $I_{DC}$ for the same device and $H_x$. $H_{eff}^z$ scales linearly with $I_{DC}$. We fit the $H_{eff}^z$ vs. $I_{DC}$ with a linear function: the slope of the fitted function is plotted against $H_x$ in Fig. 2(b). Note that here we have converted the bias current $I_{DC}$ to the current density that flows through the HM layer ($J_e$). The $H_x$ at which $H_{eff}^z/J_e$ saturates corresponds to the DM exchange field $H_{DM}$[14]. We find $H_{DM}$~200 Oe for devices with $d$~0.1 nm.

The in-plane field dependence of the current induced DW velocity is also measured to estimate $H_{DM}$. The distance a DW moved when voltage pulses are applied is measured using a Kerr microscopy. The DW velocity is estimated by dividing the distance the DW traveled by the pulse length. Fig. 2(c) shows the velocity for up-down (black squares) and down-up (red circles) DWs plotted as a function of $H_x$. From these plots, $H_{DM}$ is estimated from the $H_x$ at which the velocity becomes zero. Linear lines are fitted to the data to estimate $H_{DM}$.

The DM exchange constant $D$ is calculated from $H_{DM}$ using the following relation $D = \mu_0 M_s H_{DM} \Delta$. $\Delta = \sqrt{\frac{A}{K_{eff}}}$, where $A$ is the exchange stiffness and here we used $A$ = 31 pJ/m from previous reports[26]. $M_s$ and $K_{eff}$ are obtained from interpolating the data shown in Figs. 1(c) and 1(d). The black squares (green circles) in Fig. 2(d) show $D$ estimated using the anomalous Hall shift (DW velocity). $D$ decreases from ~0.2 mJ/m² to ~0 mJ/m² with increasing Ta layer thickness.

The DW resistance is obtained by nucleating multiple DWs into the wire and measuring the wire's resistance. DWs are nucleated with a single voltage pulse (~28 V, 100 ns) applied to the wire without any magnetic field. After nucleation, $H_z$ is applied to change the number of DWs in the wire. $H_z$ is then reduced to zero and the resistance and the magnetic state of the wire are studied using transport and Kerr measurements, respectively. $|H_z|$ is subsequently increased to further change the number of DWs. This process is repeated with increasing $H_z$ to annihilate all DWs in the wire.

Figure 3(a) shows representative Kerr images of the wire when $H_z$ is varied. The associated change of the wire resistance is shown in Fig. 3(b). We record the change in the wire resistance ($\delta R$) when the number of DWs in the wire changes due to the application of $H_z$ (note that the resistance is measured at zero field). The DW resistance ($\Delta R$) is obtained by dividing $\delta R$ with the number of DWs annihilated at the corresponding field. We find little dependence of $\Delta R$ on $H_z$. Such measurement is repeated for more than 100 times to gain statistics of the DW resistance. Normalized histograms of $\Delta R$ for films with $d$~0.1, 0.5 and 0.8 nm are shown in Fig. 4(a). The histograms are fitted with a Gaussian function to find the mean value, which is plotted in Fig. 4(b) (black squares) as a function of $d$. $\Delta R$ is positive for all devices regardless of the Ta layer thickness. The value of DW resistance is ~20 mΩ.

To account for the DW resistance observed, we first examine contributions from the SMR and AMR. The magnetization rotates within the $zx$ ($zy$) plane across a Néel (Bloch) DW leading to a non-zero $x(y)$ projection of the magnetization at the DW where the AMR (SMR) contribution is relevant. As shown in Fig. 1(f), the magnitude of AMR is orders of magnitude smaller than SMR and thus we may neglect its influence on $\Delta R$. The two solid lines in Fig. 4(b) show the range of DW resistance when contributions of SMR are taken into account. The upper blue (lower red) line indicates the level of $\Delta R$ provided the DWs are Néel (Bloch) type. As evident, contribution from SMR on *ΔR* is smaller than what has been observed experimentally (Fig. 4(b), black squares) despite the relatively large SMR found in this system. Moreover the DW resistance due to SMR is negative: i.e. the uniformly magnetized state has a larger resistance compared to the state containing DWs. Thus clearly *ΔR* found in the experiments have a different origin.

Next, the DW resistance that occurs due to the anomalous Hall effect (AHE)[27-29] is considered. Electrons are deflected from the current flow direction when the AHE is present in perpendicularly magnetized systems. Similar deflections occur when perpendicular magnetic field is applied to conductors via the Hall effect. Across a domain wall, electrons will experience deflection in opposite directions, resulting in an abrupt change of their trajectory. Such an abrupt trajectory change results in additional resistance at the DW: the effect is proportional to the square of the anomalous Hall angle of the FM layer. We use finite element simulation to estimate the AHE induced DW resistance. The electric potential distribution in the wire is calculated with and without a DW. (The length and width of the wire is varied to exclude geometrical effects on the DW resistance.) The change in the wire resistance due to the presence of a DW is obtained as the AHE induced DW resistance. Measured values of the *d*-dependent anomalous Hall angle ($\varphi_{\text{AHE}}$) of the CoFeB layer are used in the simulations. $\varphi_{\text{AHE}}$ is obtained from $\varphi_{\text{AHE}}^{\text{eff}}$ by excluding current shunting into other conducting layers: the resistivity of the three conducting layers (W, Ta, CoFeB) used in the calculations are ~120, ~200 and ~160 μΩ cm for W, Ta and CoFeB, respectively[25]. The calculated DW resistance due to AHE is plotted by the green circles in Fig. 4(b). *ΔR* due to AHE is positive and its magnitude is close to that of the experimental results. We find a relatively good agreement between the experiments and the combined contributions from the AHE and SMR for the thick Ta layer films where *D* is small and the DWs are Bloch-type. However, a discrepancy between the two (experiments vs. AHE and SMR) remains for the thinner Ta layer films: a negative DW resistance of the order ~10 mΩ is needed to account for the difference.

One well known DW resistance that may influence the measurements is the so-called intrinsic contribution due to spin accumulation at the DW[30,31]. The intrinsic DW resistance is expected to increase as the DW width becomes smaller. As $K_{eff}$ increases with decreasing Ta layer thickness, the DW width, which scales with $(K_{eff})^{-\frac{1}{2}}$, becomes smaller for thinner Ta layer films. As the intrinsic contribution is typically positive[32,33], we infer that this effect cannot account for the difference of $\Delta R$ between the experiments and the estimation of the thinner Ta layer films.

Recently, it has been reported that a chiral DW resistance appears in system with large spin orbit coupling[23]. Provided a Rashba-like Hamiltonian can describe the electronic states of the heterostructure, an additional DW resistance emerges only when the magnetic configuration is a chiral Néel type. A theoretical model predicts that the DW resistance is negative for chiral magnetic structures[23]. We therefore infer that the chiral DW resistance may partly account for the difference between the experiments and the estimation for the thin Ta layer films. According to Ref. [23] the chiral DW resistance is expressed as $R_{chi} \sim (\pi R_{DW} \Delta D)/(2A)$[23,34]. We estimate $R_{chi} \sim 2$ mΩ for chiral Néel DWs if we take $R_{DW}$ from the thicker Ta films wherein no chiral DW resistance is assumed. As a few approximations are made in the formula above, further investigation is required to clarify contribution from the chiral DW resistance.

In summary, we have studied the DW resistance in W/Ta/CoFeB/MgO heterostructures as a function of Ta insertion layer thickness. The thickness of the Ta layer influences the DW type (Néel-like walls or Bloch walls) via changes in the Dzyaloshinskii Moriya interaction. In constrast, we find the DW resistance shows little variation with the Ta layer thickness. The DW resistance of the thick Ta layer films can be accounted for by taking into consideration the SMR and AHE of the heterostructures. However, a discrepancy remains for the thin Ta layer films, wherein the DWs are chiral Néel-like DWs. We infer that chiral DW resistance, or other unknown source, contributes to the resistance in such systems. These results show that, despite the large SMR, DW resistance measurements cannot be used to identify the type of DWs.

## Acknowledgements

The authors thank Daichi Chiba and Kyung-Jin Lee for discussion on domain wall magnetoresistance. This work was partly supported by JSPS Grant-in-Aids (15H05702, 16H03853), Casio Foundation, and Center of the Spintronics Research Network of Japan. Y.-C.L. is a JSPS international research fellow.

**Figure captions**

**Fig. 1.** (a) Schematic illustration of the film stacking. The film structure is sub.|3 W|*d* Ta|1 CoFeB|2 MgO|1 Ta (units in nanometer). (b) Optical micrographs of a representative wire and Hall bar with the definition of the coordinate axis. (c-f) Saturation magnetization per unit volume $M_S$ (c), effective magnetic anisotropy energy density $K_{eff}$ (d), anomalous Hall angle $\varphi_{AHE}^{eff}$ (e) and the spin Hall magnetoresistance (red circles) and the anisotropic magnetoresistance (blue squares) (f) plotted as a function of the Ta layer thickness *d*.

**Fig. 2.** (a) Anomalous Hall loops for two different dc currents ($I_{DC} \sim \pm 6$mA) for a sample with $d \sim 0.1$ nm. The bias field along x, $H_x$ is fixed to $\sim -1100$ Oe. Definition of $H_{eff}^z$ is schematically illustrated. (b) Inset: $H_{eff}^z$ vs. $I_{DC}$ for the same sample as in (a) with $H_x \sim \pm 1100$ Oe. A linear function is fitted to data. The slope of the linear function divided by the current density ($J_e$) that flows in the HM layer is plotted against $H_x$ in the main panel. $H_{DM}$ is extracted from this plot as schematically drawn. ($H_{DM} \sim 200$ Oe here). (c) DW velocity as a function of $H_x$ for a sample with $d \sim 0.1$ nm. Data are fitted with a linear function and $H_{DM}$ is estimated from the field at which the velocity is zero. (d) The DM exchange constant *D* plotted as a function of Ta layer thickness *d*. Black squares and green circles represent *D* obtained using the anomalous Hall loop shift measurements and the DW velocity measurements, respectively.

**Fig. 3.** (a) Kerr microscopy images of the magnetic states after applying an out-of-plane magnetic field $H_z$ indicated in the left. The bright and dark regions correspond to magnetization pointing along +z and –z, respectively (see Fig. 1(b) for the definition of the coordinate axis).

(b) Number of domain walls (black squares, left axis) and the corresponding resistance of the wire (blue circles, right axis) plotted as a function of $H_z$.

**Fig. 4.** (a) Normalized histograms of DW resistance (per one DW) $\Delta R$. From the top, $d$~0.8nm, ~0.5nm and ~0.1nm. The number of DW resistance measurements is more than 700 for each wire. (b) Black squares show the measured DW resistance as a function of Ta layer thickness $d$. The error bars correspond to standard deviation of the histograms. The green circles represent calculated $\Delta R$ due to contribution from the anomalous Hall effect. The broken line shows the range of $\Delta R$ when AMR and SMR contributions are taken into account: the upper blue and lower red limit correspond to $\Delta R$ when the DW is Néel and Bloch walls, respectively.

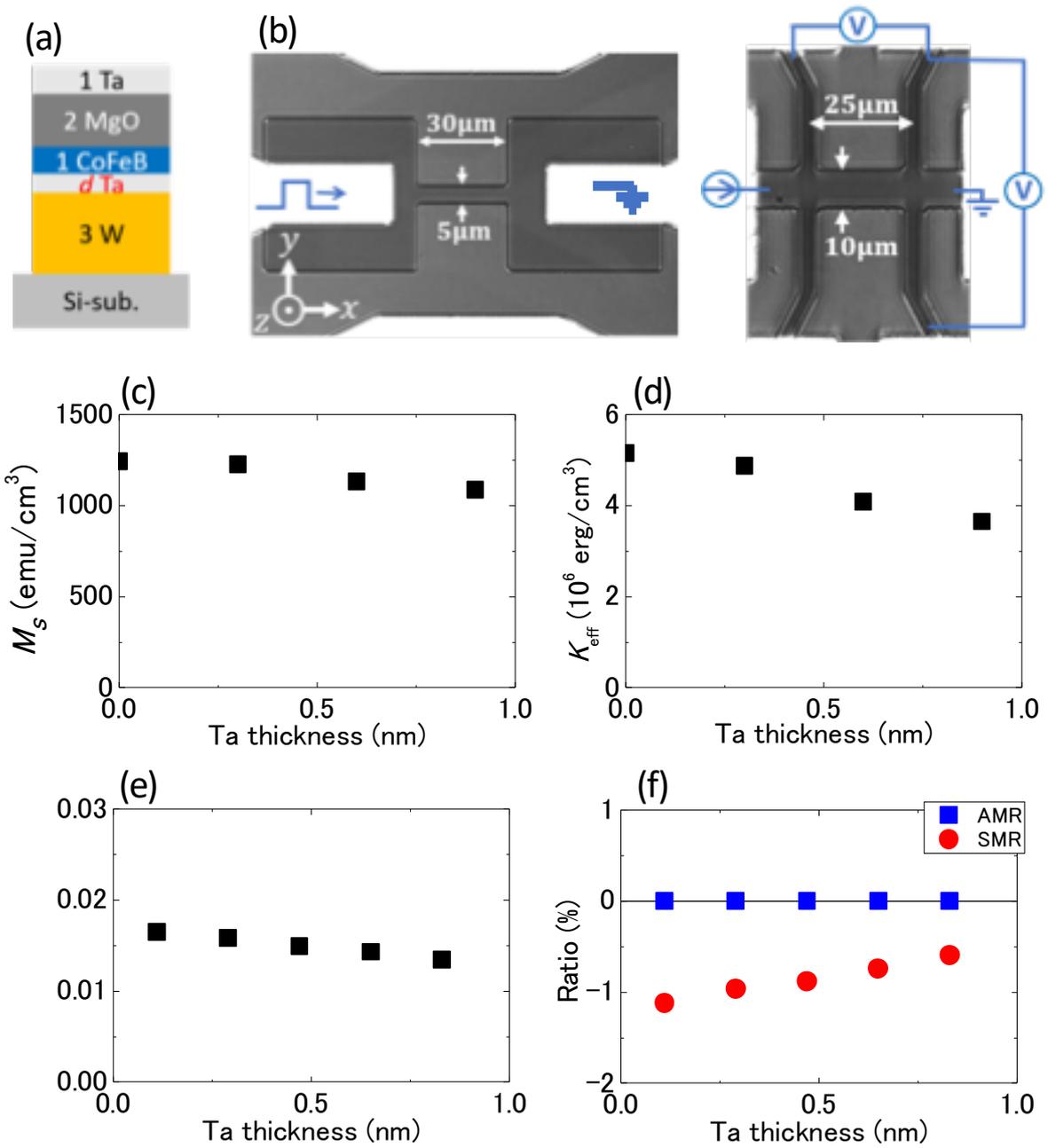

Fig. 1

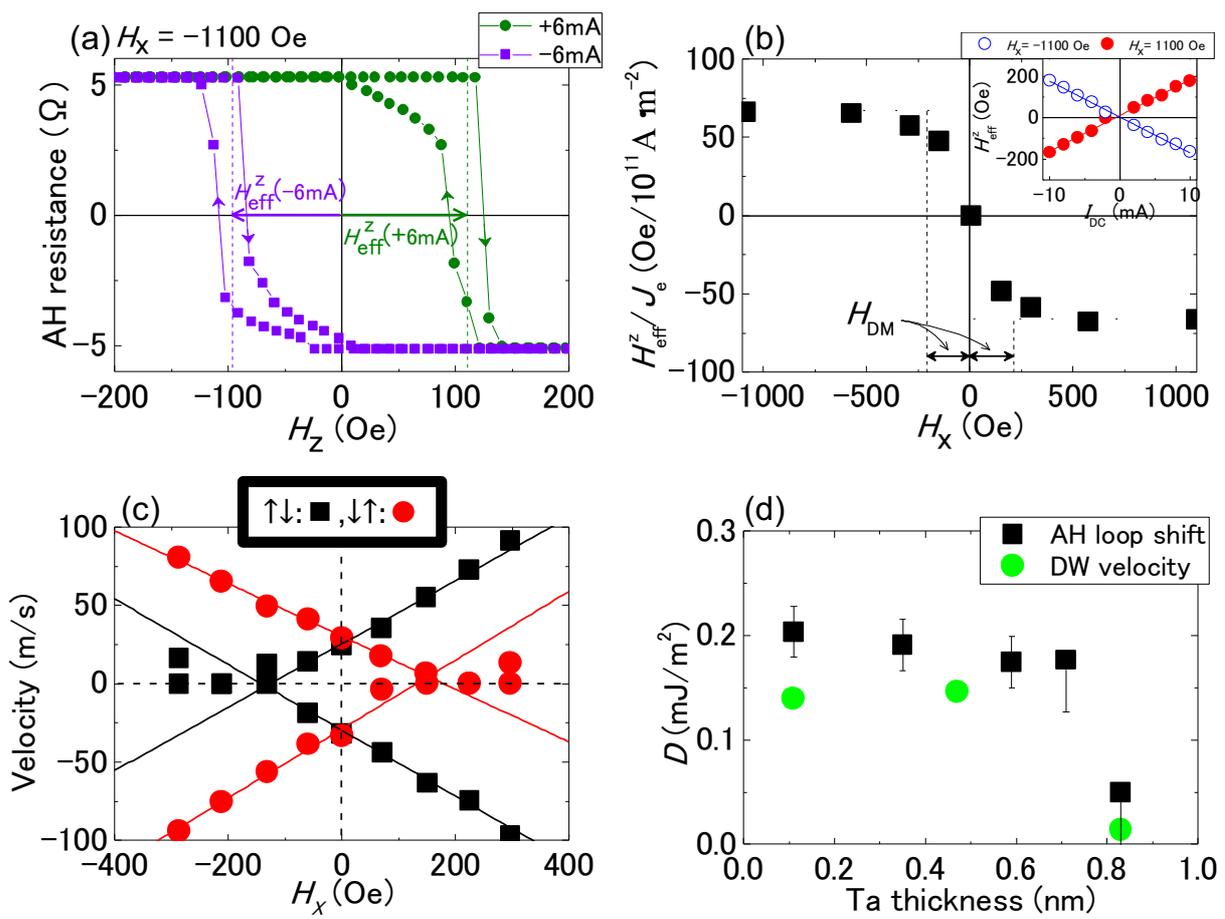

Fig. 2

(a) 3 W|**0.7** Ta|1 CoFeB|2 MgO

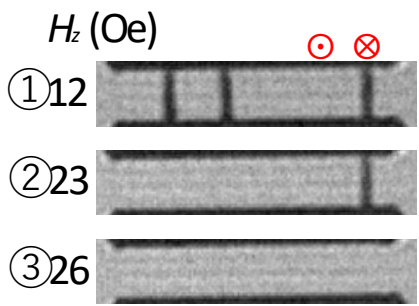

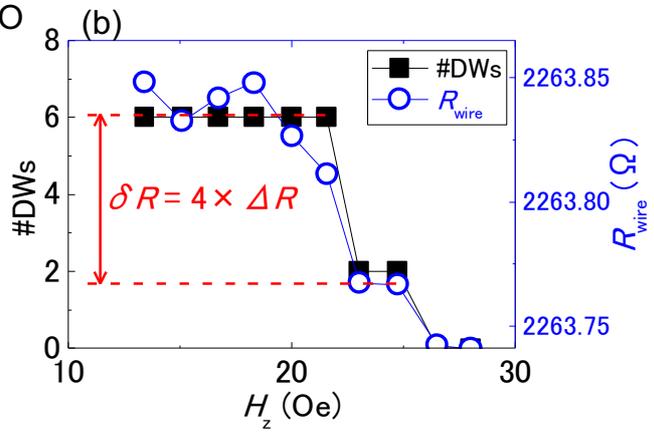

Fig. 3

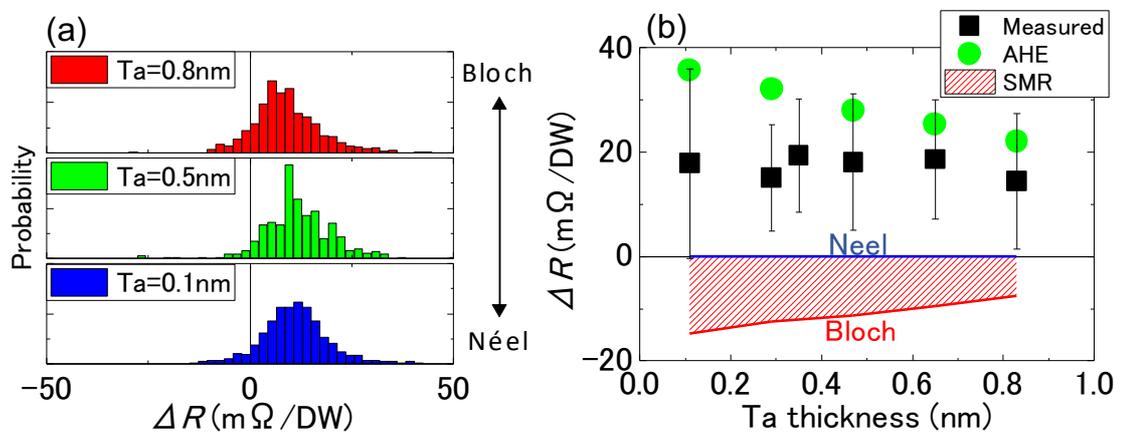

Fig. 4